%% file: main.tex
\documentclass[sigconf,nonacm]{acmart}

\renewcommand\footnotetextcopyrightpermission[1]{} 

\AtBeginDocument{%
  }

\setcopyright{acmlicensed}
\copyrightyear{2018}
\acmYear{2018}
\acmDOI{XXXXXXX.XXXXXXX}
\acmConference[Conference acronym 'XX]{Make sure to enter the correct
  conference title from your rights confirmation email}{June 03--05,
  2018}{Woodstock, NY}
\acmISBN{978-1-4503-XXXX-X/2018/06}

\usepackage{amsmath,amsfonts}
\usepackage{algorithmic}
\usepackage[ruled,vlined,linesnumbered]{algorithm2e}
\usepackage{graphicx}
\usepackage{textcomp}
\usepackage{cleveref}
\usepackage{amsmath}
\usepackage{subfig}
\usepackage{xcolor}
\usepackage{soul}
\usepackage{listings}
\usepackage{tikz}
\usepackage{bm}
\usepackage{xspace}

\usepackage{sc25repro}

\input{macros}

\begin{document}

\title{Task-Based Programming for Adaptive Mesh Refinement in Compressible Flow Simulations}

\author{Anjiang Wei}
\affiliation{%
  \institution{Stanford University}
  \city{Stanford}
  \state{CA}
  \country{USA}
}
\email{anjiang@cs.stanford.edu}  

\author{Hang Song}
\affiliation{%
  \institution{Stanford University}
  \city{Stanford}
  \state{CA}
  \country{USA}
}
\email{songhang@stanford.edu}  

\author{Mert Hidayetoglu}
\affiliation{%
  \institution{Stanford University}
  \city{Stanford}
  \state{CA}
  \country{USA}
}
\email{merth@cs.stanford.edu}  

\author{Elliott Slaughter}
\affiliation{%
  \institution{SLAC National Accelerator Laboratory}
  \city{Menlo Park}
  \state{CA}
  \country{USA}
}
\email{eslaught@slac.stanford.edu}  

\author{Sanjiva K. Lele}
\affiliation{%
  \institution{Stanford University}
  \city{Stanford}
  \state{CA}
  \country{USA}
}
\email{lele@stanford.edu}  

\author{Alex Aiken}
\affiliation{%
  \institution{Stanford University}
  \city{Stanford}
  \state{CA}
  \country{USA}
}
\email{aiken@cs.stanford.edu}  

\begin{abstract}
High-order solvers for compressible flows are vital in scientific applications. Adaptive mesh refinement (AMR) is a key technique for reducing computational cost by concentrating resolution in regions of interest. In this work, we develop an AMR-based numerical solver using Regent, a high-level programming language for the Legion programming model. We address several challenges associated with implementing AMR in Regent. These include dynamic data structures for patch refinement/coarsening, mesh validity enforcement, and reducing task launch overhead via task fusion. Experimental results show that task fusion achieves 18× speedup, while automated GPU kernel generation via simple annotations yields 9.7× speedup for the targeted kernel. We demonstrate our approach through simulations of two canonical compressible flow problems governed by the Euler equations.
\end{abstract}





\maketitle

\input{1_intro}
\input{3_math}
\input{4_design}
\input{5_perf}
\input{6_demo}

\input{2_related}
\input{7_future}
\input{8_conclusion}


\bibliographystyle{ACM-Reference-Format}
\bibliography{references}


\end{document}

%% file: macros.tex
\Crefname{equation}{Eq.}{Eqs.}
\Crefname{figure}{Fig.}{Figs.}

\newcommand{\CodeIn}[1]{\texttt{#1}}

\newcommand{\inlinespeedup}{18×\xspace}

\newcommand{\gpuspeedup}{9.7×\xspace}

\lstdefinestyle{legion}{
  language=C,
  basicstyle=\ttfamily\scriptsize,  
  keywordstyle=\color{blue!70!black}\bfseries,
  commentstyle=\itshape\color{gray!65},
  frame=lines,
  keepspaces=true,
  columns=fullflexible,
  breaklines=true,
  breakindent=0pt,
  morekeywords={fspace,ispace,region,int1d,int3d, bool}
}

\lstdefinestyle{regent}{
  language=C,
  basicstyle=\ttfamily\scriptsize,  
  keywordstyle=\color{blue!70!black}\bfseries,
  commentstyle=\itshape\color{gray!65},
  frame=lines,
  keepspaces=true,
  columns=fullflexible,
  breaklines=true,
  breakindent=0pt,
  morekeywords={reads, writes, where, do, task, end, region, ispace}
}

\definecolor{Green}{rgb}{0.6,1,0.8}

%% file: 1_intro.tex
\section{Introduction}
\label{sec:intro}

In simulations of fluid flows, adaptive mesh refinement (AMR) offers significant computational savings by concentrating mesh resolution on fine-scale features that must be accurately resolved or captured, while coarsening the mesh in regions of no interest~\cite{berger1984adaptive, berger1989local}. In computational fluid dynamics (CFD), this capability makes AMR particularly effective for simulating compressible flows, multiphase flows, and related phenomena, where structures such as shock waves and interfaces are present and interact with the flow motion~\cite{garnier2009large, fuster2009numerical}. Large-scale simulations with AMR require a robust and efficient infrastructure. Several open-source AMR frameworks, such as Chombo~\cite{colella2009chombo}, AMReX~\cite{zhang2019amrex}, SAMRAI~\cite{gunney2016advances}, and Enzo~\cite{bryan2014enzo}, have been developed using the traditional MPI+X programming model and are widely used in scientific computing. This work demonstrates the use of a task-based parallel programming model to develop an AMR-based numerical solver for compressible flow simulations.

Task-based programming systems~\cite{slaughter2015regent,bauer2012legion,augonnet2009starpu,chamberlain2007parallel} have recently emerged as a performant and productive paradigm for programming modern supercomputers. These systems model computations as user-defined \emph{tasks} and lift data into abstract representations that enable automated analysis and optimization. In contrast to the MPI+X programming model, task-based systems allow users to express applications at a higher level of abstraction. Regent~\cite{slaughter2015regent}, the framework adopted in this work, is a high-level programming language for the Legion programming model~\cite{bauer2012legion}, further streamlining application development and performance tuning.

Several challenges arise in this setting. First, the dynamic nature of adaptive mesh refinement complicates the design of data structures for AMR-based numerical solvers. Since patches, which are portions of the mesh assigned to individual processors, can be dynamically allocated (refined) and deallocated (coarsened) at runtime rather than defined statically, the data structures must flexibly represent the parent and child relationships among patches. Second, this dynamic behavior necessitates well-defined rules for refinement and coarsening to ensure the validity of the mesh structure and the correctness of the computations. Finally, task-based systems pose unique performance challenges, particularly the runtime overhead incurred by frequent task launches~\cite{slaughter2020task}.

Our work addresses the aforementioned challenges. We design data structures within the Legion programming model that support the dynamic behavior of adaptive mesh refinement. We also establish and formalize the rules required to ensure correctness and present algorithms that enforce these constraints during runtime. Finally, we observe that in task-based systems, overly fine-grained tasks can incur substantial runtime overhead, a challenge that has been systematically analyzed and quantified in prior work~\cite{slaughter2020task}. Regent provides a mechanism to mitigate this overhead through automatic task fusion~\cite{sundram2022task,yadav2025composing}, enabled by simple compiler directives, eliminating the need for manual code modifications.

Our results (shown in \Cref{sec:perf}) demonstrate that compiler-directed task fusion achieves up to \inlinespeedup speedup, effectively amortizing the runtime overhead associated with task-based systems. Moreover, Regent, as a high-level programming language, enables GPU offloading through simple compiler annotations, significantly reducing the effort required to develop high-performance kernels. This automated GPU code generation yields up to \gpuspeedup speedup compared to CPU execution for the targeted kernel.

We present the physical and mathematical formulations of the application in \Cref{sec:math}. The numerical solver targets the compressible Euler equations under the assumption of a calorically perfect gas, and implements a Riemann solver algorithm~\cite{shu1988efficient}. Demonstrative simulations of two canonical compressible flow problems involving inviscid calorically perfect gas are shown in \Cref{sec:demo}.

In summary, this paper makes the following contributions:
\begin{itemize}
\item \textbf{Design.} We design and implement a numerical solver within a task-based programming model that supports dynamic refinement and coarsening.
\item \textbf{Performance.} We demonstrate that Regent, a high-level programming language, enables effective performance tuning through simple annotations, achieving up to \inlinespeedup speedup via task fusion and up to \gpuspeedup speedup through automated GPU kernel generation.
\item \textbf{Application.} We present the physical and mathematical formulations of the compressible Euler equations and showcase demonstrative simulations of two canonical compressible flow problems.
\end{itemize}

%% file: 3_math.tex
\section{Physical and Mathematical Formulations}
\label{sec:math}

The solver numerically solves the compressible Euler equations assuming calorically perfect gas. The physical and mathematical models are described in \Cref{sec:equations}, and the elementary numerical schemes used in the solver and detailed solution process in each time advancement is illustrated in \Cref{sec:schemes}.

\subsection{Governing Equations}\label{sec:equations}
The compressible Euler equations contain the conservation of mass, momentum, and total energy.
\begin{align}
    \frac{\partial\rho}{\partial t} + \frac{\partial\rho u_j}{\partial x_j} &= 0\label{eqn:cons_mass}
    \\
    \frac{\partial\rho u_i}{\partial t} + \frac{\partial}{\partial x_j}\left(\rho u_i u_j + p\delta_{ij}\right) &= 0\label{eqn:cons_mmt}
    \\
    \frac{\partial\rho e}{\partial t} + \frac{\partial}{\partial x_j}\left(\rho h u_j\right) &= 0\label{eqn:cons_enrg}
\end{align}
The equations are expressed using the Einstein index notation following the summation convention, where $t$ and $x_j$ represent the time and space coordinate vector, respectively; $\rho$ is density; $u_i$ is the velocity vector; $p$ is pressure; $\delta_{ij}$ is the identity tensor; $e$ is the specific total energy; and $h$ is the specific total enthalpy. The specific total energy includes two components representing the specific internal energy, $e_\mathrm{th}$, and kinetic energy, $u_ju_j/2$, respectively.
\begin{equation}
    e = e_\mathrm{th} + u_ju_j/2
\end{equation}
The specific total enthalpy, $h$, is calculated as
\begin{equation}
    h = e + p / \rho
\end{equation}
The equation system is closed by the equation-of-state (EOS) model of a calorically perfect gas parameterized by a specific gas constant, $R$, and the ratio of specific heats, $\gamma$. $R$ is not explicitly involved in the equation system since the temperature is not included in the formulation. According to the EOS model, the internal energy, density, and pressure yields the following relation.
\begin{equation}
    p = \rho e_\mathrm{th}(\gamma - 1)
\end{equation}
The speed of sound, $c$, which is used in the numerical solver to determine the time advancement step size and the approximate Riemann solver, is calculated consistent with the calorically perfect gas EOS model.
\begin{equation}
    c = \sqrt{\gamma p / \rho}
\end{equation}

\subsection{Basic Numerical Schemes and Solution Process}\label{sec:schemes}
\begin{figure}[!ht]
    \centering
    \scalebox{0.55}{\input{figures/UniformMesh1D.tikz}}
    \caption{Schematic of a uniform computational mesh in 1D.}
    \label{fig:mesh1d}
\end{figure}
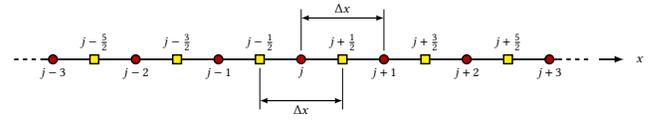

The numerical solution approach yields high-order finite difference methods. The representation of continuous solution profiles and differential operations are conducted on a discretized space-time domain. The spatial discretization forms a computational mesh. A schematic of spatial discretization of a 1D computational mesh is shown in \Cref{fig:mesh1d}, where the spatial coordinate, $x$, is discretized uniformly in to the nodal points, $x_i = x_0 + i\Delta x$, and edge points, $x_{i\pm1/2} = x_i \pm \Delta x/2$, for $i\in\mathbb{N}$. For the discretized quantities, the subscript only indicates the discrete location.

The overall solution process follows the simulation framework proposed in prior work~\cite{song2022robust, song2024robust}, where the conservative variables are stored and advanced at collocated nodal points while the fluxes are assembled at the staggered edge points in each dimension via the interpolation of a minimum complete set of primitive variables, $p$, $T$, and $u_j$. Other primitive variables needed for the flux assembly are locally computed at the staggered edge points. The time advancement is conducted using the 3-stage strong-stability-preserving Runge-Kutta (SSP-RK3) method~\cite{shu1988efficient}. As demonstrated in prior work~\cite{song2022robust, song2024robust, song2024numerical}, the numerical framework results in significantly reduced aliasing error and effectively enhances numerical robustness while preserving high spectral resolution.

For simulations of shock-free flows, 4th-order central mid-point interpolation and staggered central difference schemes are applied for flux assembly and divergence, respectively. For a variable $\phi$, the schemes are expressed in the following two equations.
\begin{align}
    &\phi_i
    = \frac{9}{8}\frac{\phi_{i+\frac{1}{2}} + \phi_{i-\frac{1}{2}}}{2} - \frac{1}{8}\frac{\phi_{i+\frac{3}{2}} + \phi_{i-\frac{3}{2}}}{2} + \mathcal{O}(\Delta x^4)
    \\
    &\left.\frac{\partial\phi}{\partial x}\right|_{x_i}
    = \frac{9}{8}\frac{\phi_{i+\frac{1}{2}} - \phi_{i-\frac{1}{2}}}{\Delta x} - \frac{1}{8}\frac{\phi_{i+\frac{3}{2}} - \phi_{i-\frac{3}{2}}}{3\Delta x} + \mathcal{O}(\Delta x^4)
\end{align}
For flows that contain shocks, nonlinear interpolation using a 5-point weighted essentially non-oscillatory (WENO5-JS) scheme is applied~\cite{jiang1996efficient, shu2009high}. The nonlinear interpolation scheme obtains the result as a convex superposition of the results calculated using a set of candidate stencils. The weights for the superposition are determined based on the solution smoothness within each candidate stencil. The stencil that contains a non-smooth solution will be given less or zero weight. If the solution is smooth over all candidate stencils, the superposition will reach the 5th-order linear interpolation scheme. This approach has been widely applied for shock-capturing in high-order numerical simulations~\cite{shu2009high}. Using the nonlinear interpolation, the fluxes are assembled using the Rusanov method~\cite{rusanov1962calculation, toro2013riemann} as an approximate Riemann solver. The Euler equations, \Cref{eqn:cons_mass} -- \Cref{eqn:cons_enrg}, can be denoted in a matrix-vector form.
\begin{equation}
    \frac{\partial\bm{U}}{\partial t} + \frac{\partial\bm{F}_j}{\partial x_j} = \bm{0}
    \label{eqn:euler_system}
\end{equation}
where the vector $\bm{U}$ represents the conservative variables, and the matrix $\bm{F}_j$ represents the fluxes along the dimension of $x_j$. The terminologies of ``matrix'' and ``vector'' in this context refer to the mathematical structure only and do not imply any physical meaning. According to the Rusanov approach, the Riemann fluxes, $\bm{F}_j^{\mathrm{Riemann}}$, are computed as follows.
\begin{equation}
    \bm{F}_j^{\mathrm{Riemann}}
    = \frac{1}{2}\left(\bm{F}_j^R + \bm{F}_j^L\right) - \frac{1}{2}S_j\left(\bm{U}^R - \bm{U}^L\right)
    \label{eqn:riemann}
\end{equation}
where $\bm{U}^R$ and $\bm{U}^L$ are the conservative variables at the staggered edge points calculated from ``left'' and ``right'' biased interpolated characteristic variables, respectively. The calculation of the characteristic variables in each dimension is detailed in Ref.~\cite{song2024robust}. Correspondingly, $\bm{F}_j^R = \bm{F}_j(\bm{U}^R)$ and $\bm{F}_j^L = \bm{F}_j(\bm{U}^L)$ are fluxes assembled using left- and right- interpolated conservative variables respectively. For simplicity, $S_j$ in \Cref{eqn:riemann} is a characteristic wave speed which is calculated as
\begin{equation}
    S_j = \max\left(c^R + \left|u_j^R\right|, c^L + \left|u_j^L\right|\right)
\end{equation}
where $c^R$, $c^L$, $u_j^R$, and $u_j^L$ are the speed of sound and velocity component in the $x_j$ dimension calculated using the right- and left-biased interpolated characteristic variables, respectively. In the simulation, $\bm{F}_j^\mathrm{Riemann}$ will be used to calculate the divergence-of-flux in \Cref{eqn:euler_system}.

%% file: figures/UniformMesh1D.tikz
\begin{tikzpicture}
    \draw[line width=1.2pt] (-6.2,0) -- (6.2,0);
    \draw[line width=1.2pt, dashed] (-6.2,0) -- ++(-0.8,0);
    \draw[line width=1.2pt, dashed] ( 6.2,0) -- ++( 0.8,0);
    \draw[line width=1.2pt, -latex] (7.2, 0) -- (7.8,0);
    \node[anchor=west] at (8,0) {$x$};
    
    \draw[line width=0.8pt, fill=black!30!red] (0,0) circle (3pt);
    \node[anchor=north, yshift=-2pt] at (0, 0) {$j$};
    \foreach \i in {1,...,3} {
        \draw[line width=1.0pt, fill=black!30!red] ( 2*\i,0) circle (3pt);
        \draw[line width=1.0pt, fill=black!30!red] (-2*\i,0) circle (3pt);
        \node[anchor=north, yshift=-3pt] at ( 2*\i, 0) {$j+\i$};
        \node[anchor=north, yshift=-3pt] at (-2*\i, 0) {$j-\i$};
    }
    \foreach \i in {1,3,5} {
        \draw[line width=1.0pt, fill=yellow, xshift=-3pt, yshift=-3pt] ( \i, 0) rectangle ++(6pt, 6pt); 
        \draw[line width=1.0pt, fill=yellow, xshift=-3pt, yshift=-3pt] (-\i, 0) rectangle ++(6pt, 6pt); 
        \node[anchor=south, yshift=3pt] at ( \i, 0) {$j+\frac{\i}{2}$};
        \node[anchor=south, yshift=3pt] at (-\i, 0) {$j-\frac{\i}{2}$};
    }
    
    \draw[line width=0.5pt, yshift=6pt] (0,0) --++ (0,1);
    \draw[line width=0.5pt, yshift=6pt] (2,0) --++ (0,1);
    \draw[line width=0.8pt, latex-latex] (0,1) --++ (2,0) node[anchor=south] at (1,1) {$\Delta x$};
    
    \draw[line width=0.5pt, yshift=-6pt] (-1,0) --++ (0,-1);
    \draw[line width=0.5pt, yshift=-6pt] ( 1,0) --++ (0,-1);
    \draw[line width=0.8pt, latex-latex] (-1,-1) --++ (2,0) node[anchor=north] at (0,-1) {$\Delta x$};
\end{tikzpicture}

%% file: 4_design.tex
\section{Solver Design and Implementation}
\label{sec:design}

\subsection{Legion Programming Model}

\emph{Tasks} are specialized functions that can be scheduled in parallel across a distributed system. They do not rely on global variables or external references tied to a single address space, ensuring portability and scalability. Tasks may have multiple \emph{variants} optimized for different processor types, such as separate implementations for CPUs and GPUs.

In Legion, tasks are issued to the runtime in program order but can be reordered or executed in parallel, provided that sequential semantics are preserved. The Legion runtime handles dependence analysis~\cite{bauer2023visibility} to enforce correct execution~\cite{lee2018correctness}. Like functions, tasks can invoke other tasks, referred to as \emph{parent} and \emph{child} tasks, respectively.

\emph{Index launches}~\cite{soi2021index} enable the execution of multiple parallel tasks in a single bulk operation. Each task in an index launch is identified by a multi-dimensional \emph{launch domain}, which defines the set of task instances.

\emph{Regions} are data collections that can be partitioned and distributed across a machine. Each region consists of an $n$-dimensional index space and a field space, similar to a struct. Conceptually, regions generalize multi-dimensional arrays and are analogous to tensors (e.g., in TensorFlow~\cite{abadi2016tensorflow}), distributed datasets (e.g., in Spark~\cite{zaharia2010spark}), and arrays (e.g., in Chapel~\cite{chamberlain2007parallel}). Tasks explicitly declare the regions they access, specifying whether they read from, write to, or apply reductions to them.

\subsection{Data Structure Design}
\label{subsec:data-structure}

\begin{figure}[t]
\centering
\begin{lstlisting}[style=legion]
/* -------- field space -------- */
fspace grid_meta_fsp {
 level      : int,      /* AMR refinement level */

 nbr        : int[8],   /* 0-3: N, E, S, W  (direct neighbours); 4-7: NE, SE, SW, NW (diagonals); value -1 means physical boundary */
 parent     : int,      /* coarser patch PID */
 child      : int[4],   /* finer  patch PIDs */
 refine_req : bool,     /* flag: request refinement */
 coarsen_req: bool      /* flag: request coarsening */
}

/* -------- region instantiation (excerpt) -------- */
np_max = grid.num_patches_max;
fp_sz  = grid.full_patch_size;
ng     = grid.num_ghosts;

color_space = ispace(int1d, np_max, 0);

patches_grid = region(ispace(int3d,
                      {np_max, fp_sz, fp_sz},
                      {0, -ng, -ng}),
                      grid_fsp);

patches_meta = region(ispace(int1d, np_max, 0),
                      grid_meta_fsp);
\end{lstlisting}
\caption{Patch metadata and region allocation.}
\label{fig:grid-meta}
\end{figure}

\begin{figure}[!htb]
    \centering
    \scalebox{0.9}{\input{figures/patches.tikz}}
    \caption{Index space of the actual data patches.}
    \label{fig:patch_schematics}
\end{figure}
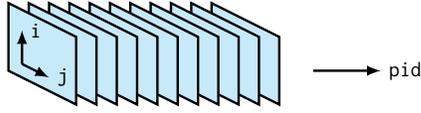

The elementary unit of the data structure is the patch. Patches are the building blocks of the computational grid. All patches have the same grid size. Each patch is identified by a unique
\(\mathtt{pid}\). During the runtime, patches are dynamically assigned to the spatially hierarchical computational domain based on the needs of local grid refinement and coarsening.
The patch is a logical concept that contains two components -- metadata and actual data. The metadata refers to the information of the current patch relative to the entire computational domain, such as the active status, relative position, parent patch, and child patch IDs, if any. The actual data refers to the values that represent a discretized quantity stored at the computational grid points. The metadata and actual data are constructed in two individual Legion regions and are invertibly linked. Numerical operations are applied to the active data patches, and the issuing is guided by the metadata of the corresponding patch. The metadata and actual data patches are collected into different regions, which are defined in the Legion programming model~\cite{bauer2012legion}. The region consists of an index space (\texttt{ispace}) and field space (\texttt{fspace}). Data collected in a region is accessed by a combination of an index and the name of the field. The metadata and actual data regions can be expressed as follows:
\begin{itemize}
    \item \textbf{Metadata}: $\mathtt{ispace}(\mathtt{pid})\otimes\mathtt{fspace}(\text{metadata contents})$
    \item \textbf{Actual data}: $\mathtt{ispace}(\mathtt{pid}, i, j)\otimes\mathtt{fspace}(\text{variables})$
\end{itemize}
The index space of the metadata only contains the dimension of $\mathtt{pid}$, while the index space of the region of actual data patches contains both $\mathtt{pid}$ and the local coordinates, $i$ and $j$, as shown in \Cref{fig:patch_schematics}. The index space that defines the size of each patch spans both the interior grid points and the ghost grid points for the halo communication. The field space of the region of data patches contains the physical variables, such as the physical coordinates, conservative variables, and buffer primitive variables. The field space of the region of metadata are shown in \Cref{fig:grid-meta}.

If a patch is active, the global index, corresponding to the computational grid, can be calculated using the metadata and the local coordinates as the offset. Each patch have at most 8 neighboring patches, including four sides and four corners. The neighboring patches for halo communications are identified by the field \texttt{nbr}. The across-resolution topology is managed by the \texttt{parent} and \texttt{child} fields. The adaptive grid refinement and coarsening during runtime are controlled by the fields \texttt{refine\_req} and \texttt{coarsen\_req}, respectively. These two flags are set by the runtime program as requests. During runtime, only legal requests that do not conflict with the basic refinement and coarsening rules will be processed. Further details regarding the adaptive refinement and coarsening are illustrated in \Cref{sec:towards_amr}.

\subsection{Towards Adaptive Mesh Refinement}\label{sec:towards_amr}

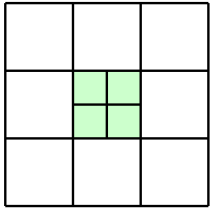
\begin{figure}[!htb]
    \centering
    \scalebox{0.9}{\input{figures/RefineValid.tikz}}
    \caption{A valid multi-resolution mesh pattern.}
    \label{fig:multi-res_mesh}
\end{figure}

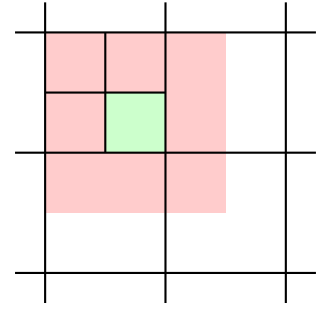
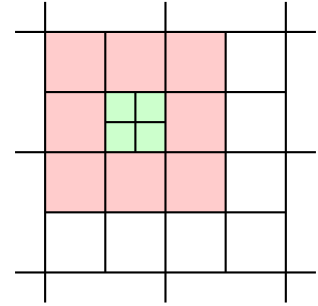
\begin{figure}[!htb]
    \centering
    \subfloat[]{\scalebox{0.8}{\input{figures/RefineStep1.tikz}}\label{fig:refine1}}\quad
    \subfloat[]{\scalebox{0.8}{\input{figures/RefineStep2.tikz}}\label{fig:refine2}}
    \caption{Illustration of the refinement process. The refinement request is marked on the green patch, and its Moore neighborhood is highlighted in red. \protect\subref{fig:refine1} shows the original local mesh structure, and \protect\subref{fig:refine2} shows the updated mesh structure after refinement.}
    \label{fig:refinement}
\end{figure}

\SetKwComment{Comment}{$\triangleright$\ }{}
\SetKwFunction{RefinePatch}{RefinePatch}
\SetKwProg{Fn}{Function}{:}{}
\SetFuncSty{textbf}  
\SetCommentSty{textit}  

\begin{algorithm}
\caption{Recursive Refinement Procedure}
\label{algo:refine_patch}

\Fn{\RefinePatch{$\mathcal{M}$, $p$}}{
  \Comment{$\mathcal{M}$ is the set of active patches}
  \Comment{$p$ is the patch to refine}

  \If{$p$ is a leaf}{
    \ForEach{$q \in \mathcal{N}(p)$}{ \Comment{check every Moore neighborhood position}
      \If{$q$ does not exist}{
        $r \gets p.\text{parent.corresponding\_neighbor}$\;
        \RefinePatch{$\mathcal{M}$, $r$}\;
      }
    }
    allocate $p.\text{children}$\;
    set meta-data for $p.\text{children}$\;
    update overall mesh structure $\mathcal{M}$
  }
}
\end{algorithm}

\begin{algorithm}
\caption{Adaptive Mesh Coarsening}
\label{algo:amr_coarsen}

\SetKwFunction{CoarseningPass}{CoarseningPass}
\SetKwFunction{IsCoarseningAllowed}{IsCoarseningAllowed}
\SetKwFunction{DeleteChildren}{DeleteChildren}

\SetFuncSty{bfseries}

\SetCommentSty{bfseries}
\SetCommentSty{myCommentFont}

\newcommand{\myCommentFont}[1]{\textit{\textcolor{gray}{#1}}}

\SetKwProg{Fn}{Function}{:}{}
\SetKwComment{Comment}{$\triangleright$\ }{}

\Fn{\CoarseningPass{$\mathcal{M}$}}{
  \ForEach{patch $p \in \mathcal{M}$ flagged for coarsening}{
    \If{\IsCoarseningAllowed{$p, \mathcal{M}$}}{
      \DeleteChildren{$p, \mathcal{M}$}\;
    }
  }
}

\vspace{0.5em}
\Comment{Helper functions for coarsening}

\Fn{\IsCoarseningAllowed{$p, \mathcal{M}$}}{
  \If{$p.\text{level} = 0$ \textbf{or not} $p.\text{has\_children}$}{ \KwRet \textbf{false} }
  \ForEach{neighbor $q \in \mathcal{N}(p)$}{
    \If{$q$ exists \textbf{and} $q.\text{has\_children}$}{ \KwRet \textbf{false} }
  }
  \KwRet \textbf{true}\;
}

\Fn{\DeleteChildren{$p, \mathcal{M}$}}{
  DeallocatePatches($p.\text{child\_ids}$)\;
  $p.\text{child\_ids} \gets \text{null}$\;
}
\end{algorithm}

The base-level patches form a Cartesian layout. For a multi-resolution mesh, each base-level patch serves as the root node of a quad tree in 2D. To support the numerical methods used in the computation, a valid multi-resolution mesh structure must satisfy the following condition -- for any patch, its parent patch, if it exists, must fully contain the corresponding Moore neighborhood, as illustrated in \Cref{fig:multi-res_mesh}. A boundary patch can also meet this condition by introducing ghost patches, which contain the ghost grid points to enforce boundary conditions. If a domain is periodic, then all boundary patches are treated equivalently as interior patches.

The solution-dependent multi-resolution mesh adaptation process will support the accessibility of the complete stencil required by the finite difference methods. To achieve this, two rules must be enforced during the refinement and coarsening processes. Let $P$ be an active patch and let $\mathcal N(P)=\{P_i\}_{i=1}^{8}$ denote its Moore neighborhood, including four face-adjacent and four corner-adjacent patches. The two rules, R1 and R2, are described as follows.

\noindent\textbf{R1: Persistence of the base mesh:}

All root-level patches (level 0) form a permanent coarsest scaffold of the
domain throughout the lifetime of a computation job. Consistent with the restriction, the root-level patches can never be removed during coarsening processes, and a root-level patch with no child patch characterizes the ultimate termination condition for mesh coarsening.

\noindent\textbf{R2: Recursive refinement and coarsening.}

For any active patch $P$ on a level $l$, if $P$ has child patches on the level $l+1$, then $\mathcal{N}(P)$ must be active. This rule is necessary due to the multi-resolution numerical methods employed in the simulation. Specifically, if $l=0$, $P$ is on the root level. Consistent with R1, a complete set of $\mathcal{N}(P)$ is guaranteed to exist. This indicates that the mesh refinement from level $0$ to level $1$ does not have any further constraints. However, for $l\ge1$, the refinement and coarsening processes must be aware of this rule. For a refinement request marked for a patch $P$ on level $l$, if R2 is not satisfied, a pre-processing step is needed to ensure the presence of all patches that belong to $\mathcal{N}(P)$. A schematic is shown in \Cref{fig:refinement}. This pre-processing step may invoke recursion, depending on the dynamic structure of the multi-resolution mesh. \Cref{algo:refine_patch} presents the implementation of the recursive refinement procedure.

For a coarsening request marked for a patch $P$ on level $l$, if the children of $P$ are not leaf patches, then the coarsening request is invalid. Additionally, if any of the child patches of $P$ serve as the Moore neighborhood to support a valid mesh structure on a higher level than $l$, the coarsening request will not be processed. Otherwise, the request is accepted directly, the child patches are deleted, and $P$ becomes a leaf patch. The coarsening procedure is presented in \Cref{algo:amr_coarsen}.

%% file: figures/patches.tikz
\begin{tikzpicture}
\foreach \x in {0,...,10} {
    \draw[line width=1pt, fill=cyan!20!white] (-0.3*\x, 0) --++ (1, -0.5) --++ (0, 1) --++ (-1, 0.5) -- cycle;
}
\draw[line width=1.0pt, -latex] (1.5, 0) --++ (1, 0) node[anchor=west] {\texttt{pid}};
\draw[line width=1.0pt, -latex] (-2.8, 0.1) --++ (0, 0.5) node[anchor=west] {\texttt{i}};
\draw[line width=1.0pt, -latex] (-2.8, 0.1) --++ (0.4, -0.2) node[anchor=west] {\texttt{j}};
\end{tikzpicture}

%% file: figures/RefineValid.tikz
\begin{tikzpicture}
    \draw[draw=none, fill=green!20!white] (-0.5, -0.5) rectangle (0.5, 0.5);
    \foreach \i in {-1.5, -0.5, 0.5, 1.5}{
        \draw[line width=1.0pt] (-1.5, \i) --++ (3, 0);
        \draw[line width=1.0pt] (\i, -1.5) --++ (0, 3);
    }
    \draw[line width=1.0pt] (-0.5, 0) --++ (1, 0);
    \draw[line width=1.0pt] (0, -0.5) --++ (0, 1);
\end{tikzpicture}

%% file: figures/RefineStep1.tikz
\tikzset{
    refine 1/.pic={
        \draw[line width=1.0pt] (-1, 0) --++ (2, 0);
        \draw[line width=1.0pt] (0, -1) --++ (0, 2);
    },
    basic mesh/.pic={
        \draw[draw=none, fill=red!20!white] (-2, -1) rectangle (1, 2);
        \draw[draw=none, fill=green!20!white] (-1, 0) rectangle (0, 1);
        \foreach \i in {-2,0,2} {
            \draw[line width=1.0pt] (-2.5, \i) --++ (5.0, 0);
            \draw[line width=1.0pt] (\i, -2.5) --++ (0, 5.0);
        }
        \pic at (-1, 1) {refine 1};
    },
}
\begin{tikzpicture}
    \pic at (0, 0) {basic mesh};
\end{tikzpicture}

%% file: figures/RefineStep2.tikz
\tikzset{
    refine 1/.pic={
        \draw[line width=1.0pt] (-1, 0) --++ (2, 0);
        \draw[line width=1.0pt] (0, -1) --++ (0, 2);
    },
    refine 2/.pic={
        \draw[line width=1.0pt] (-0.5, 0) --++ (1, 0);
        \draw[line width=1.0pt] (0, -0.5) --++ (0, 1);
    },
    basic mesh/.pic={
        \draw[draw=none, fill=red!20!white] (-2, -1) rectangle (1, 2);
        \draw[draw=none, fill=green!20!white] (-1, 0) rectangle (0, 1);
        \foreach \i in {-2,0,2} {
            \draw[line width=1.0pt] (-2.5, \i) --++ (5.0, 0);
            \draw[line width=1.0pt] (\i, -2.5) --++ (0, 5.0);
        }
        \pic at (-1, 1) {refine 1};
        \pic at (-1,-1) {refine 1};
        \pic at ( 1,-1) {refine 1};
        \pic at ( 1, 1) {refine 1};
        \pic at (-0.5, 0.5) {refine 2};
    },
}
\begin{tikzpicture}
    \pic at (0, 0) {basic mesh};
\end{tikzpicture}

%% file: 5_perf.tex
\section{Performance Results}
\label{sec:perf}

\begin{figure*}[!tb]
    \centering
    \begin{minipage}[b]{0.48\textwidth}
        \centering
        \includegraphics[width=\textwidth]{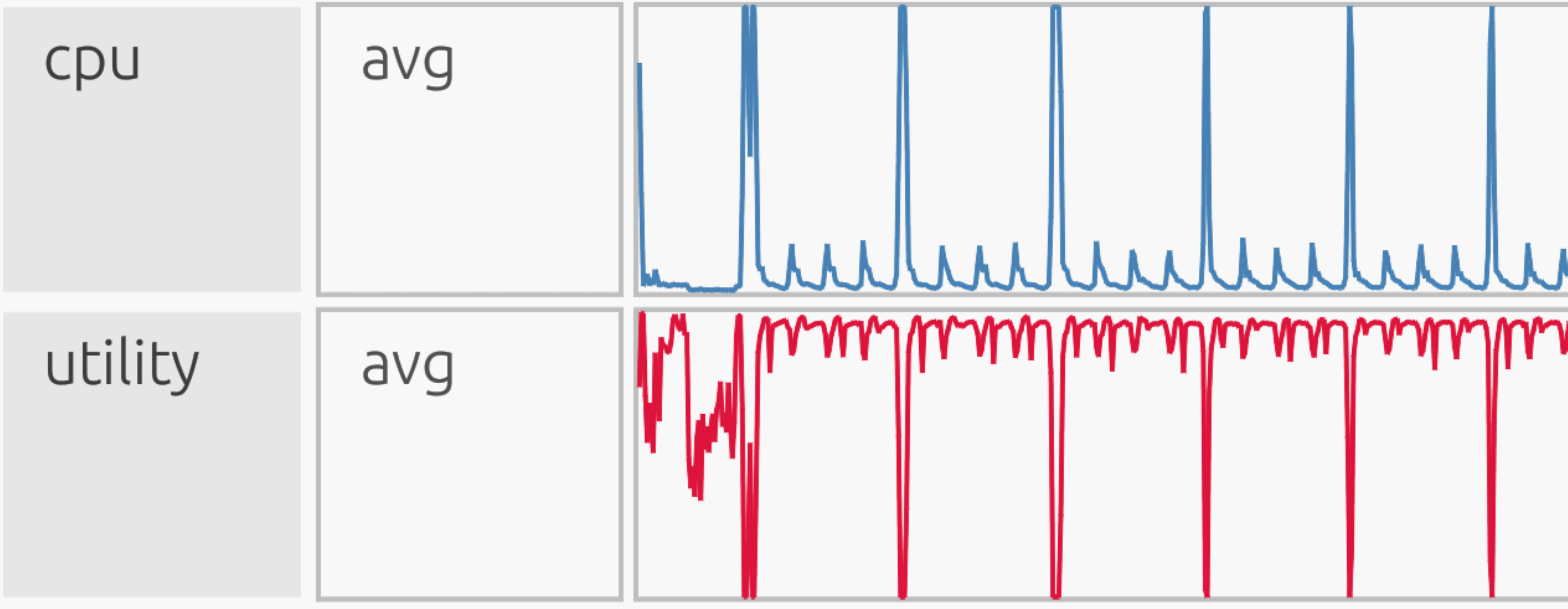}
        \caption{This profiling figure highlights the runtime overhead, as evidenced by the significant utilization of \emph{utility} processors. These processors are dedicated cores specifically allocated for the runtime system.}
        \label{fig:overhead}
    \end{minipage}
    \hfill
    \begin{minipage}[b]{0.48\textwidth}
        \centering
        \includegraphics[width=\textwidth]{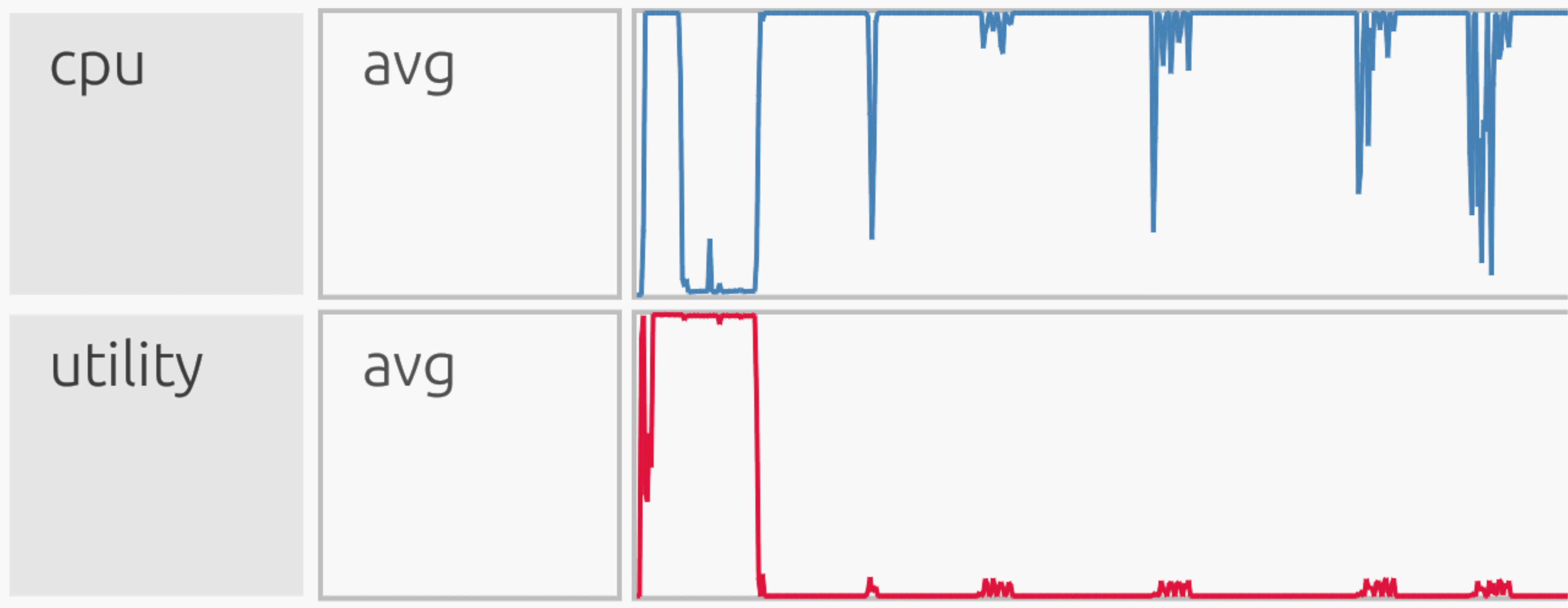}
        \caption{This profiling figure illustrates the impact of task fusion, which reduces the usage of utility processors. As a result, the CPU cores responsible for task execution become the new performance bottleneck, as expected.}
        \label{fig:inlined}
    \end{minipage}
\end{figure*}

\begin{figure}[t]
    \begin{lstlisting}[style=regent,escapeinside={(*@}{@*)},columns=fixed,xleftmargin=3.5ex]
(*@\sethlcolor{Green}\hl{+$\,$ \_\_demand(\_\_inline)}@*)
task solver.calcGVC(
    gvc : region(ispace(int3d), GRAD_VEL),
    cvn : region(ispace(int3d), CVARS),
    mp  : region(ispace(int1d), grid_meta_fsp)
)
where
    writes(gvc),
    reads(cvn.{mass, mmtx, mmty, enrg}, mp.{level})
do
    for cij in gvc.ispace do
        ...
    end
end
\end{lstlisting}
    \caption{Illustration of the \texttt{solver.calcGVC} task with the \texttt{\_\_demand(\_\_inline)} directive highlighted, indicating the inlining demand for task fusion to mitigate runtime overhead.}
    \label{fig:inliningcode}
\end{figure}

\begin{figure}[t]
    \begin{lstlisting}[style=regent,escapeinside={(*@}{@*)},columns=fixed,xleftmargin=3.5ex]
(*@\sethlcolor{Green}\hl{+$\,$ \_\_demand(\_\_CUDA)}@*)
task ssprk3Stage(
    dt : double,
    u0 : region(ispace(int3d), CVARS),
    u1 : region(ispace(int3d), CVARS),
    u2 : region(ispace(int3d), CVARS)
)
where
    reads(u0, u1, u2),
    writes(u0)
do
    for cij in u0.ispace do
        u0[cij].mass *= ...
    end
end
\end{lstlisting}
    \caption{Illustration of the \texttt{ssprk3Stage} task with the \texttt{\_\_demand(\_\_CUDA)} directive highlighted, instructing the compiler to automatically generate GPU code.}
    \label{fig:demandcuda}
\end{figure}

We discuss the unique performance optimization challenge in task-based programming systems and demonstrate how it can be addressed with our approach in \Cref{subsec:fusion}.

\subsection{Reducing Runtime Overhead via Task Fusion}
\label{subsec:fusion}

Task-based runtimes incur non-negligible overheads for each task launch, including analysis, resource allocation, and scheduling. When these overheads are comparable to or exceed the task's execution time, they cannot be effectively amortized, resulting in significant performance degradation. In our setting, each kernel operates exclusively on patch-local data. When the patch size is small, the corresponding task executes extremely quickly, causing the runtime overhead to dominate overall performance.

As shown in \Cref{fig:overhead}, which presents the profiling results, the utility processor, responsible for runtime analysis and scheduling (red lines), exhibits high average utilization. In contrast, the CPU cores responsible for actual task execution (blue lines) remain underutilized for most of the time.

The root cause of the high utility processor usage is the small task granularity, as identified in prior work~\cite{slaughter2020task}. When tasks are too fine-grained, the overhead of launching each task dominates execution. The standard solution to this issue is task fusion~\cite{sundram2022task,yadav2025composing}, which increases task granularity so that the execution time becomes comparable to or exceeds the task launch overhead, thereby amortizing runtime costs.

Regent~\cite{slaughter2015regent}, a high-level programming language for task-based programming, eliminates the need for manual task fusion through application-level code rewriting. Instead, the issue can be addressed by simply adding a few annotations to the Regent compiler. As illustrated in \Cref{fig:inliningcode}, the \CodeIn{\_\_demand(\_\_inline)} annotation enforces that calls to the task \CodeIn{solver.calcGVC} are inlined into their caller. This eliminates the need for separate task launches and avoids the associated overhead by replacing multiple task calls per patch with a single task call per patch.

After applying task fusion, the profiling result in \Cref{fig:inlined} shows that the CPU processors (responsible for task execution) become the dominant performance bottleneck, as expected. Runtime overhead is significant only during the program’s initialization and quickly diminishes to a negligible level thereafter. The profiling results indicate nearly 100\% CPU utilization, suggesting that the CPUs are almost fully engaged in application computation, with no obvious opportunities for further performance gains. This optimization improves application performance by \inlinespeedup, reducing the per-iteration execution time from 8.3 seconds to 0.45 seconds.

\subsection{Automated Code Generation for GPUs}

In Regent, users can enable GPU execution by adding simple compiler annotations. This approach significantly streamlines the development of high-performance kernels, making it accessible to domain scientists who may not be experts in GPU programming.

\Cref{fig:demandcuda} shows an example where the task \texttt{ssprk3Stage} is annotated with the \texttt{\_\_demand(\_\_CUDA)} directive. This instructs the compiler to automatically generate GPU kernels and execute the task on the GPU at runtime. However, some tasks cannot be parallelized in this way. For example, if a loop contains loop-carried dependencies, it cannot be automatically parallelized on the GPU, and the Regent compiler will report an error in such cases.

On the GPU, the task takes 35.7 ms, compared to 347.0 ms on the CPU, resulting in a \gpuspeedup speedup. These measurements are based on a grid size of 2560 × 2560. In general, larger grid sizes lead to greater performance gains on the GPU, while smaller grid sizes result in shorter task execution times and proportionally higher runtime overhead.

%% file: 6_demo.tex
\section{Demonstrative Simulations}
\label{sec:demo}

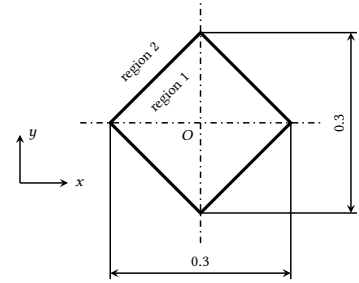
\begin{figure}[!htb]
    \centering
    \scalebox{0.8}{\input{figures/ImplosionIC.tikz}}
    \caption{Initial condition schematics of the implosion process.}
    \label{fig:implosion_ic}
\end{figure}

\begin{figure}[!htb]
    \centering
    \scalebox{0.8}{\input{figures/ShearLayerIC.tikz}}
    \caption{Schematics of the shear layer base flow configuration.}
    \label{fig:shear_layer_base}
\end{figure}
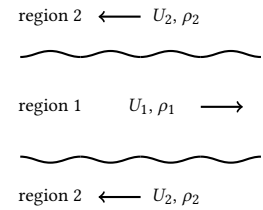

\begin{figure*}[!tb]
    \centering
    \subfloat[Implosion stage 1]{\includegraphics[width=0.40\textwidth]{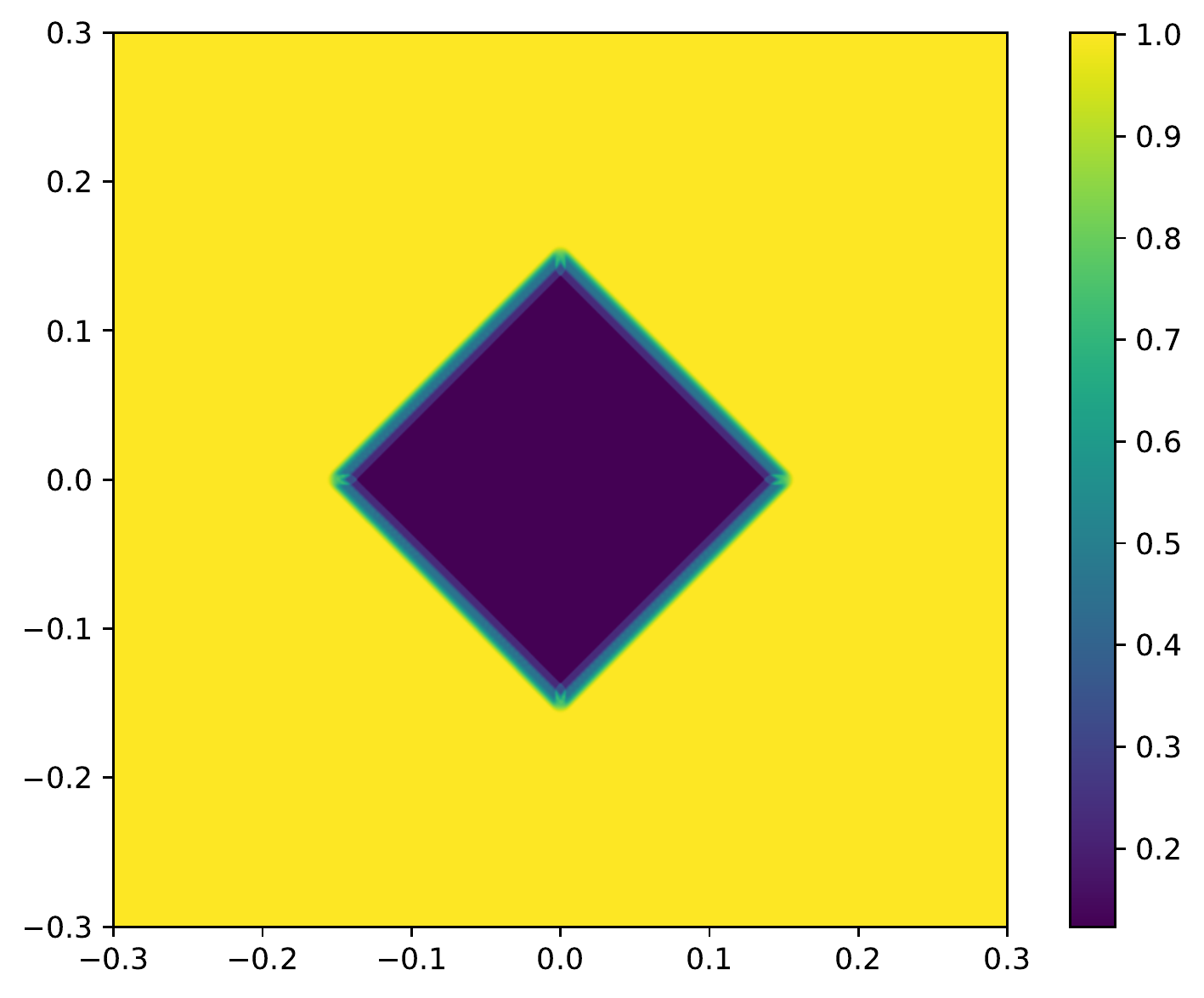}}
    \subfloat[Implosion stage 2]{\includegraphics[width=0.40\textwidth]{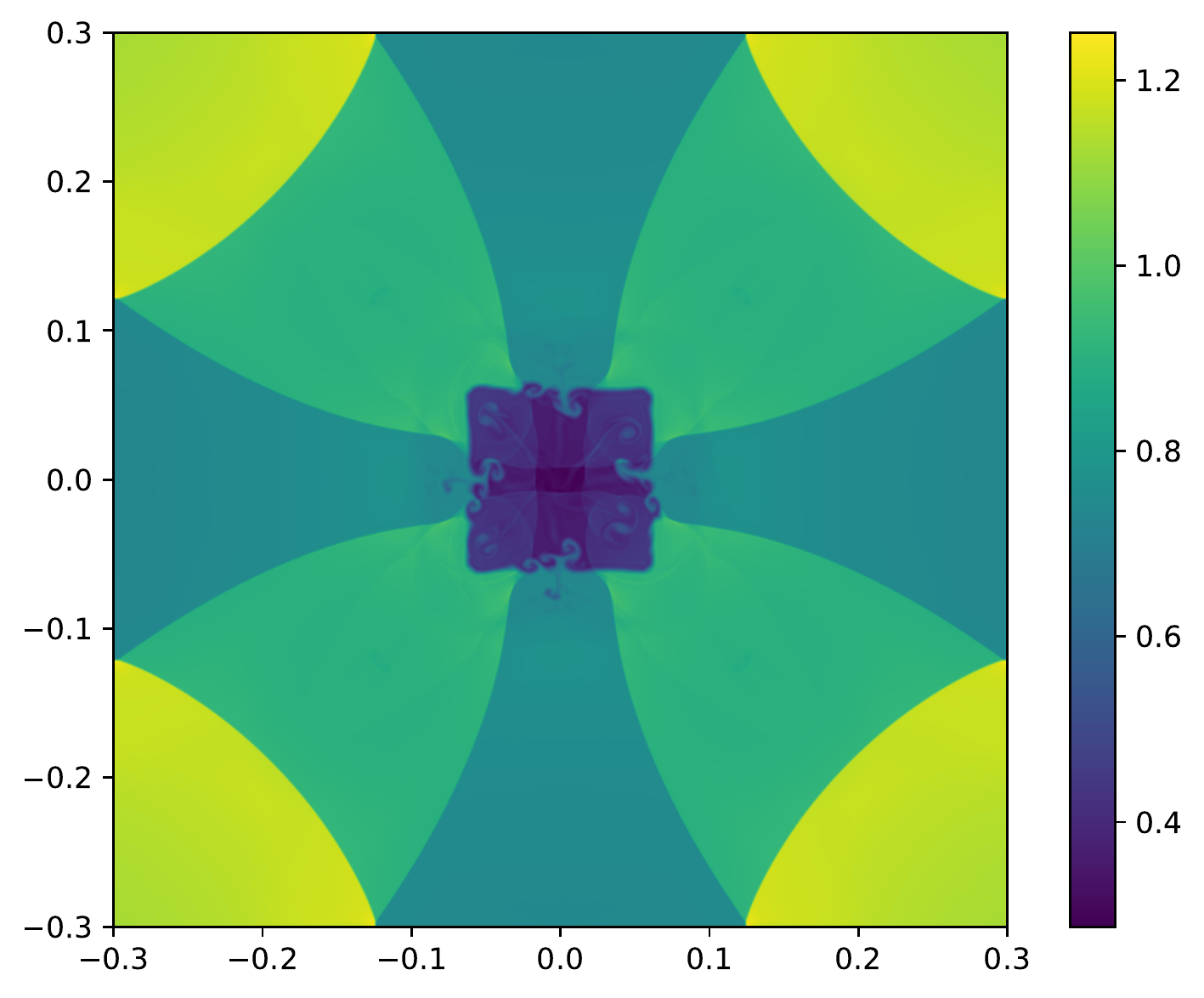}}\\
    \subfloat[Implosion stage 3]{\includegraphics[width=0.40\textwidth]{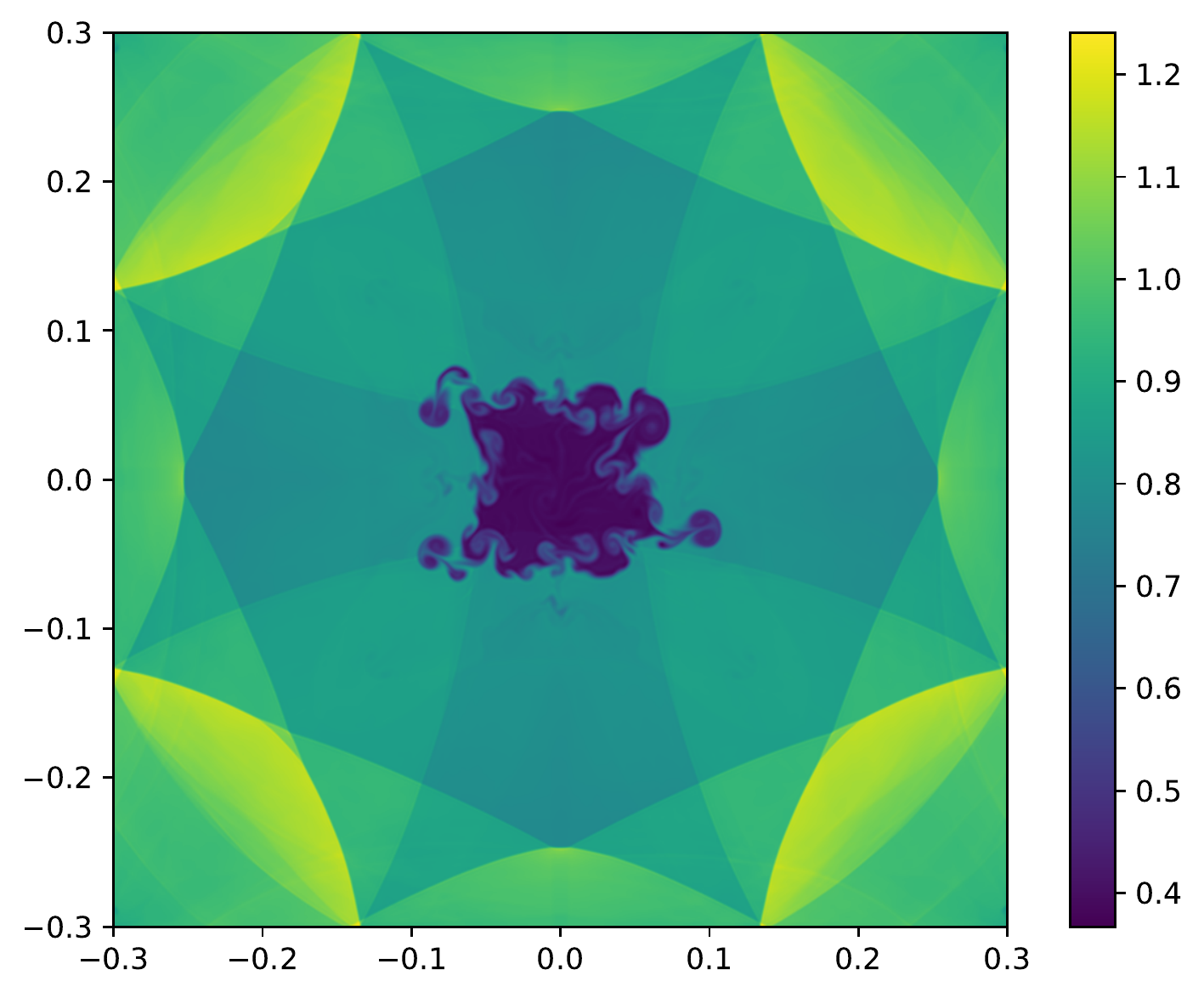}}
    \subfloat[Implosion stage 4]{\includegraphics[width=0.40\textwidth]{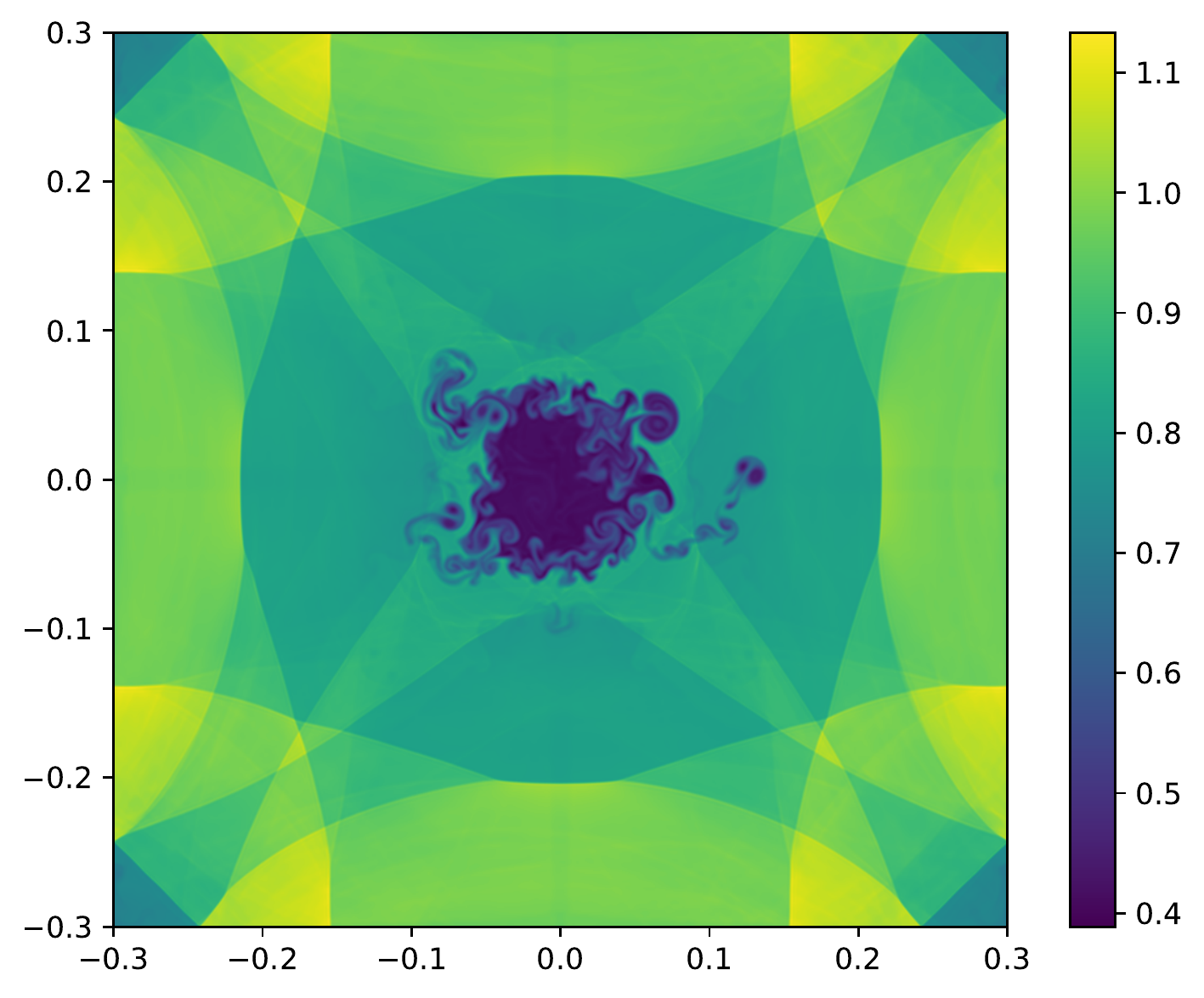}}
    \caption{Visualization of the density in the implosion process at different simulation time.}
    \label{fig:implosion}
\end{figure*}

\begin{figure*}[!tb]
    \centering
    \subfloat[Shear stage 1]{\includegraphics[width=0.40\textwidth]{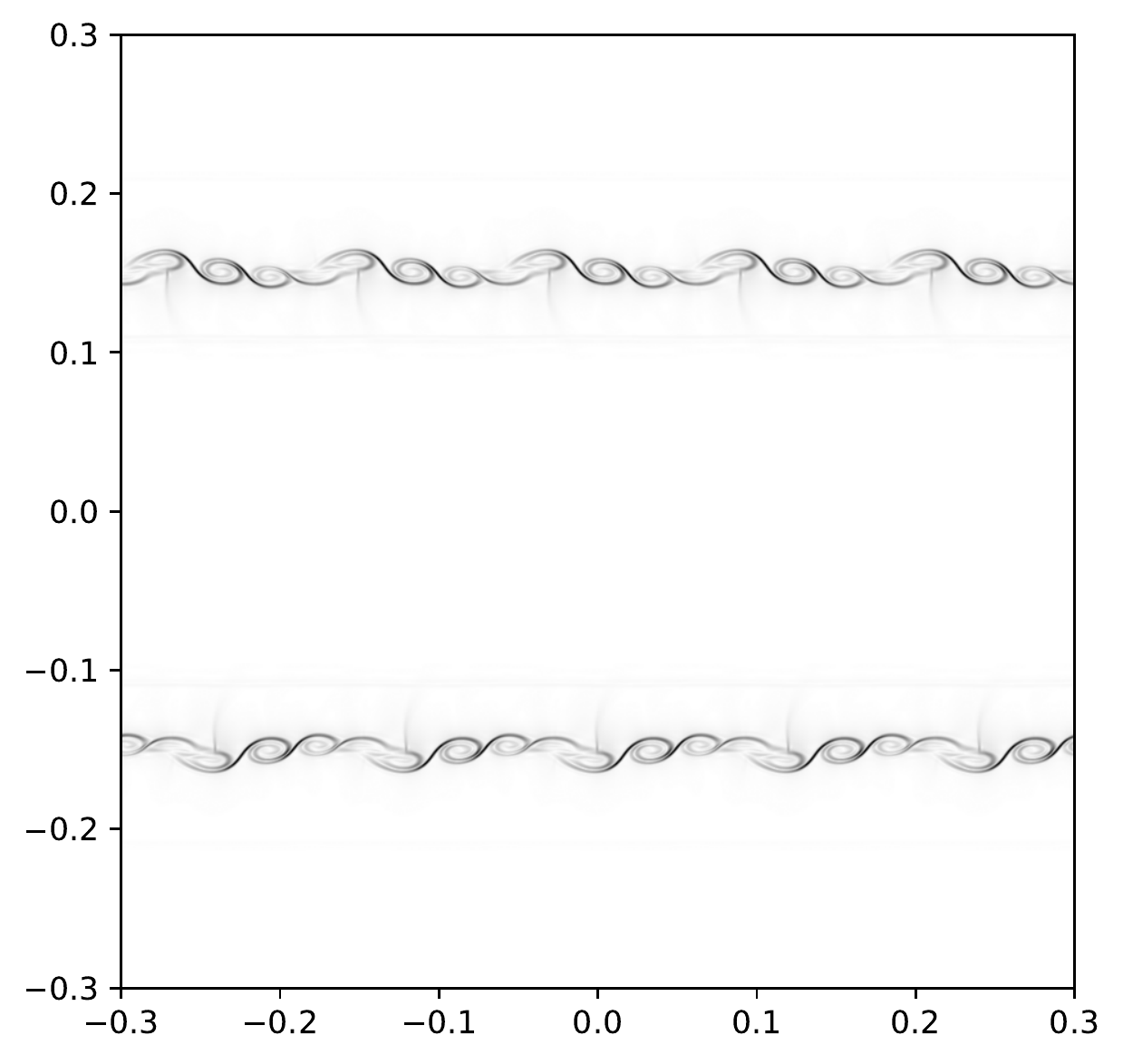}}
    \subfloat[Shear stage 2]{\includegraphics[width=0.40\textwidth]{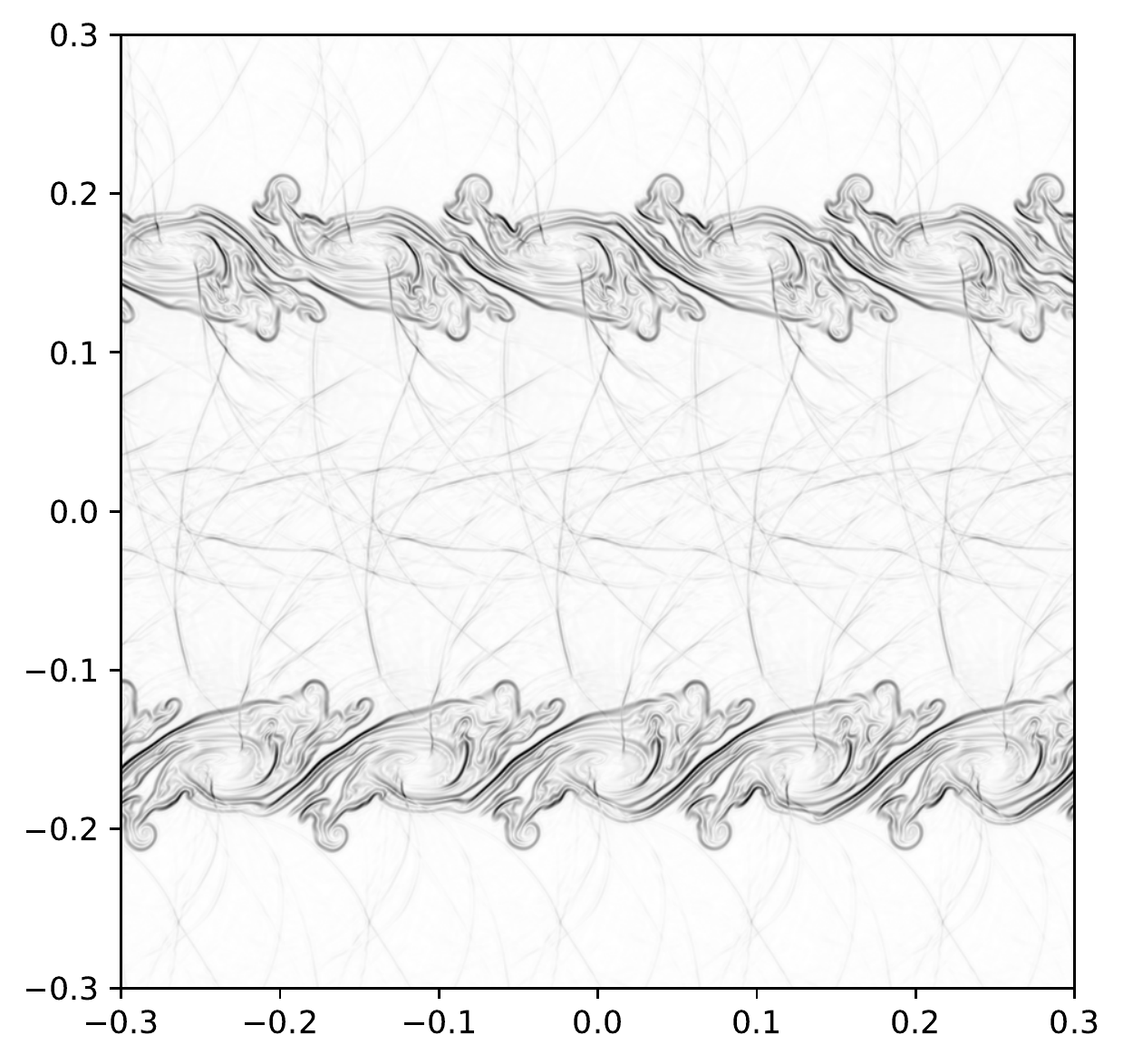}}\\
    \subfloat[Shear stage 3]{\includegraphics[width=0.40\textwidth]{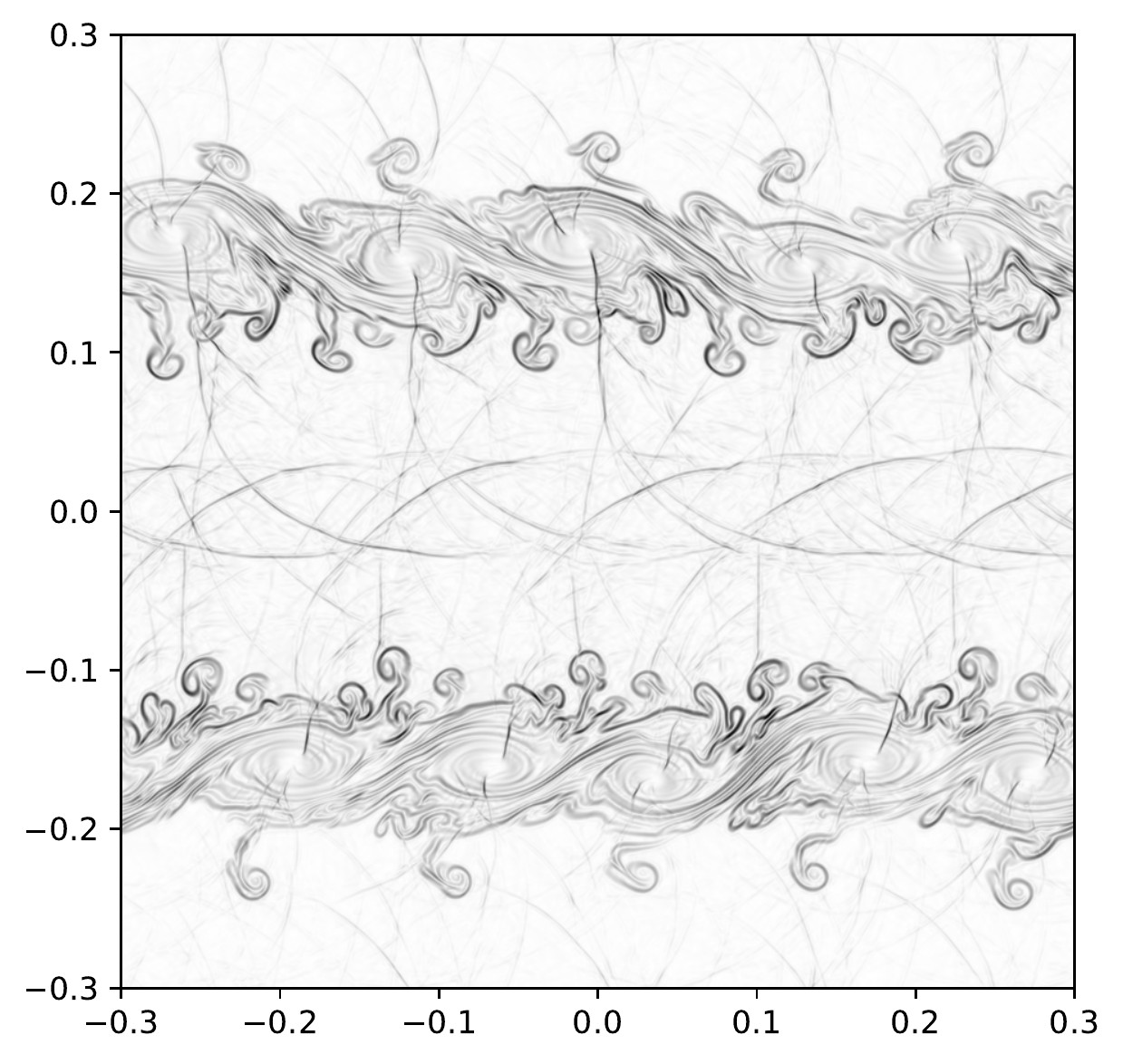}}
    \subfloat[Shear stage 4]{\includegraphics[width=0.40\textwidth]{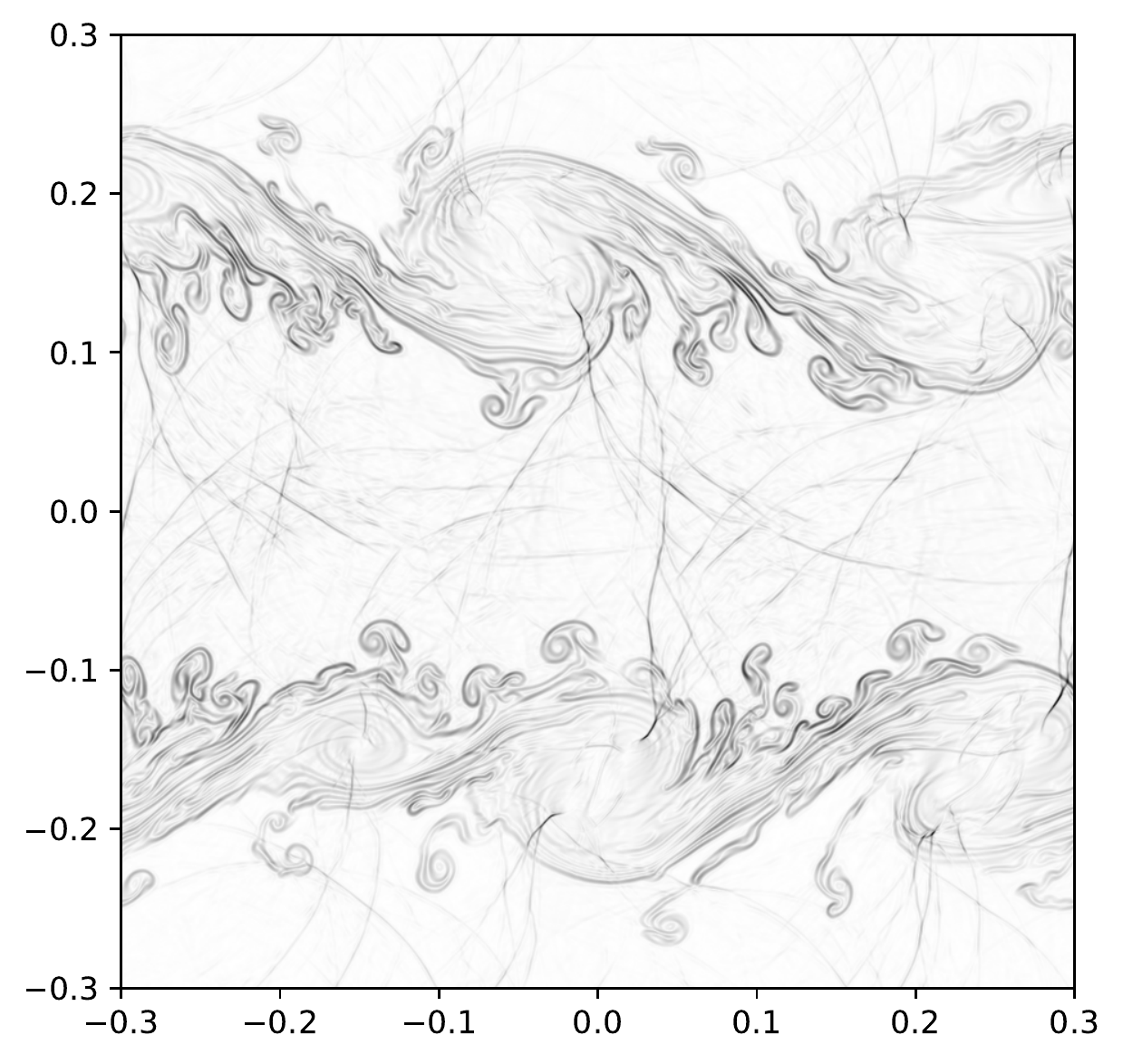}}
    \caption{Numerical Schlieren imaging of the periodic shear layer flows at different simulation time.}
    \label{fig:shear}
\end{figure*}

The solver is demonstrated using two canonical compressible flows of inviscid calorically perfect gas. The specific gas constant and ratio of specific heats are set to be $R=1$ and $\gamma=1.4$ respectively. Both demonstrative simulations are configured on 2D periodic domain and discretized on uniform Cartesian meshes. The detailed configurations and simulation results are described as follows.

The first demonstrative problem is an implosion process. The simulation is configured at $(x, y)\in(-0.3, 0.3)^2$. The flow is initially at rest. A diamond shape divides the computational domain into two regions as shown in \Cref{fig:implosion_ic}. The pressure and density are uniform in each region but sharply jump across the different regions. The initial pressure and density in each region are specified as follows.
\begin{equation}
    \left(\rho, p\right) =
    \begin{cases}
        \left(0.125, 0.140\right) & |x|+|y| <   0.15\\
        \left(1.000, 1.000\right) & |x|+|y| \ge 0.15
    \end{cases}
    \label{eqn:implosion_ic}
\end{equation}
The size of the computational mesh is $1024\times1024$, and the simulation is conducted at a CFL number equal to $0.45$. The density fields at different physical time are visualized in \Cref{fig:implosion}.

The second demonstrative problem is a compressible shear layer. The base flow is aligned in the $x$-direction. The computational domain is defined on $(x, y)\in(-0.3, 0.3)^2$. The base state is specified in two regions as shown in \Cref{fig:shear_layer_base}. Since the computational domain is periodic in both flow and transverse directions, the simulation contains two shear layers. The horizontal velocity component in the base flow is positive in region 1 and negative in region 2. The pressure in the base flow is uniform. The density is uniform within each region, but jumps across different regions. The simulation starts from the base state with a small perturbation introduced. The perturbation is localized in the transverse direction at the shear layer and periodically varies in the flow direction. The scaled wavenumber in the streamwise direction of the perturbation is $5$. The density ratio is 2.0, and the convective Mach number is 0.7. The visualization is shown in \Cref{fig:shear}.

%% file: figures/ImplosionIC.tikz
\tikzset{
    boxsty/.style={line width=1.0pt, rounded corners=3, draw=black},
    lw0/.style={line width=0.7pt},
    lw1/.style={line width=1.5pt},
}
\begin{tikzpicture}
    \draw[lw0, dashdotted] (-2,0)--++(4,0);
    \draw[lw0, dashdotted] (0,-2)--++(0,4);
    \draw[lw1] (-1.5,0)--(0,1.5)--(1.5,0)--(0,-1.5)--cycle;
    \foreach \p in {-1.5, 1.5} {
        \draw[lw0](\p,0) --++ (0,-2.6);
        \draw[lw0](0,\p) --++ (2.6, 0);
    }
    \draw[lw0, <->, >=stealth] (-1.5,-2.5)--++(3,0);
    \draw[lw0, <->, >=stealth] ( 2.5,-1.5)--++(0,3);
    \node[anchor=south] at (0,-2.5) {\footnotesize $0.3$};
    \node[anchor=south, rotate=90] at (2.5,0) {\footnotesize $0.3$};
    \node[anchor=north east] at (0,0) {\footnotesize $O$};
    \node[anchor=center, rotate=45] at (-0.5,0.5) {\footnotesize region 1};
    \node[anchor=center, rotate=45] at (-1.0,1.0) {\footnotesize region 2};
    \draw[lw0,->,>=stealth] (-3.0,-1)--++(0.8,0) node [xshift=0.2, anchor=west] {\footnotesize$x$};
    \draw[lw0,->,>=stealth] (-3.0,-1)--++(0,0.8) node [xshift=0.2, anchor=west] {\footnotesize$y$};
\end{tikzpicture}

%% file: figures/ShearLayerIC.tikz
\begin{tikzpicture}
    \foreach \x in {-2, -1, 0, 1}{
        \draw[line width=1.0pt, yshift=25pt] (\x, -0.05) to[out=0, in=180] ++ (0.5, 0.1) to[out=0, in=180] ++ (0.5, -0.1);
        \draw[line width=1.0pt, yshift=-25pt] (\x, +0.05) to[out=0, in=180] ++ (0.5,-0.1) to[out=0, in=180] ++ (0.5, +0.1);
    }
    \draw[->, line width=1.0pt] (1, 0.0) --++ ( 0.7, 0);
    \draw[->, line width=1.0pt] (0, 1.5) --++ (-0.7, 0);
    \draw[->, line width=1.0pt] (0,-1.5) --++ (-0.7, 0);
    \node at (-1.5, 0.0) {region 1};
    \node at (-1.5, 1.5) {region 2};
    \node at (-1.5,-1.5) {region 2};
    \node at ( 0.2, 0.0) {$U_1$, $\rho_1$};
    \node at ( 0.6, 1.5) {$U_2$, $\rho_2$};
    \node at ( 0.6,-1.5) {$U_2$, $\rho_2$};
\end{tikzpicture}

%% file: 2_related.tex
\section{Related Work}
\label{sec:related}

\subsection{Numerical Solvers for Compressible Fluid Dynamics}
For numerical simulations of high-speed flows, compressible Navier--Stokes equations are numerically solved with proper geometries and boundary conditions~\cite{poinsot1992boundary,pirozzoli2011numerical}. Additionally, in many aerodynamic applications where boundary layer and turbulence are neglected, the governing equations further reduces to the Euler equations. Many numerical methods have been developed for simulating compressible flows, such as the finite difference method~\cite{song2024robust}, finite volume method~\cite{eymard2000finite}, and discrete Galerkin method~\cite{luo2008discontinuous}. In order to improve the simulation accuracy and resolution with an affordable computational mesh size, high-order numerical schemes are commonly applied. Furthermore, shocks, as a discontinuity, can be developed in compressible flows~\cite{song2024numerical}. For inviscid flows, a shock is a discontinuity that represents the weak solution of the Euler system. For viscous flows, the thickness of the shock is often at a subgrid scale, which is not numerically resolved. In order to simulate flows that contain shocks, shock capturing approaches are applied. The key concept of a shock-capturing approach is to regularize a discontinuous solution to a smoothed step profile across a few grid points. There are various methods for shock, the one related to this work is the nonlinear interpolation combined with an approximate Riemann solver~\cite{shu1988efficient}.




\subsection{Parallel Programming Models}
The challenges of parallel and distributed computation have led to the development of a diverse range of programming models, languages, and runtime systems. Beyond the traditional MPI and MPI+X models, recent years have seen a proliferation of alternatives, including task-based programming systems (e.g., Legion~\cite{bauer2012legion}, Regent~\cite{slaughter2015regent}, Realm~\cite{treichler2014realm}, StarPU~\cite{augonnet2009starpu}, Sequoia~\cite{fatahalian2006sequoia}, OmpSs~\cite{duran2011ompss}, and PaRSEC~\cite{danalis2015parsec}), as well as PGAS and actor-based models (e.g., Chapel~\cite{chamberlain2007parallel}, Charm++\cite{kale1993charm++}, and X10\cite{charles2005x10}). Other notable efforts focus on portability, such as Kokkos~\cite{trott2021kokkos} and Raja~\cite{beckingsale2019raja}, while large-scale data analytics and machine learning workflows have driven the adoption of systems like Dask~\cite{rocklin2015dask}, Spark~\cite{zaharia2010spark}, TensorFlow~\cite{abadi2016tensorflow}, PyTorch~\cite{paszke2019pytorch}, Alpa~\cite{zheng2022alpa}, FlexFlow~\cite{lu2017flexflow}, and Pathways~\cite{barham2022pathways}.

Several AMR frameworks, including Chombo~\cite{colella2009chombo}, AMReX~\cite{zhang2019amrex}, SAMRAI~\cite{gunney2016advances} and Enzo~\cite{bryan2014enzo}, have been developed using the traditional MPI+X programming model, while Enzo-E~\cite{EnzoE} is an alternative built on top of Charm++. In this work, we develop our application using Regent~\cite{slaughter2015regent}, a programming language built on top of Legion~\cite{bauer2012legion}, a task-based distributed runtime system. Legion decouples performance tuning from application logic through a dedicated mapping interface~\cite{wei2025mapple}, enabling more modular optimization and simplifying performance tuning compared to traditional MPI+X models. Regent further improves programming productivity by abstracting away Legion’s low-level C++ APIs and enabling advanced compiler optimizations such as task fusion and GPU code generation, as discussed in \Cref{sec:perf}.

%% file: 7_future.tex
\section{Future Work}
\label{sec:future}

As future work, we plan to develop a more comprehensive numerical application that fully exploits the capabilities of adaptive mesh refinement. To enable efficient execution at scale, we will design an application-specific load-balancing algorithm~\cite{misaka2017adaptive} that considers both the hierarchical mesh structure and the computational heterogeneity of tasks. In addition, we will extend the system to run on multi-node heterogeneous architectures, which presents new challenges in task scheduling, minimizing inter-node communication, and efficiently moving data across CPUs and GPUs. Addressing these challenges will be essential to achieving scalable performance in more real-world scenarios.

%% file: 8_conclusion.tex
\section{Conclusion}

We have presented the design and implementation of a high-order AMR-based numerical solver for compressible flows using Regent, a high-level programming language for the Legion programming model. Our approach addresses key challenges in representing dynamic mesh structures, enforcing AMR correctness at runtime, and optimizing the performance introduced by frequent task launching overhead. Through compiler-directed task fusion and automated GPU code generation, we demonstrate that high productivity and high performance can be achieved simultaneously. Task fusion delivers up to \inlinespeedup speedup, while GPU kernel generation via simple annotations yields up to \gpuspeedup speedup for the targeted kernel. Our demonstrative simulations of canonical compressible flow problems illustrate the practical capabilities of this solver.